\documentclass[sigconf]{aamas} 

\usepackage{balance} 
\usepackage{graphicx}
\usepackage{listings}
\usepackage{float}
\usepackage{algorithm}
\usepackage{algpseudocode}

\setcopyright{ifaamas}
\acmConference[AAMAS '22]{Proc.\@ of the 21st International Conference
on Autonomous Agents and Multiagent Systems (AAMAS 2022)}{May 9--13, 2022}
{Auckland, New Zealand}{P.~Faliszewski, V.~Mascardi, C.~Pelachaud,
M.E.~Taylor (eds.)}
\copyrightyear{2022}
\acmYear{2022}
\acmDOI{}
\acmPrice{}
\acmISBN{}

\acmSubmissionID{340}

\title[Hacking the Colony]{Hacking the Colony: On the Disruptive Effect of Misleading Pheromone and How to Defend Against It}

\author{Ashay Aswale}
\authornote{All authors have contributed equally.}
\affiliation{
  \institution{Department of Robotics Engineering}
  \city{Worcester Polytechnic Institute, Worcester, MA, USA}}
\email{asaswale@wpi.edu}  
\author{Antonio López}
\authornotemark[1]
\affiliation{
  \institution{Department of Robotics Engineering}
  \city{Worcester Polytechnic Institute, Worcester, MA, USA}}
\email{alopez3@wpi.edu}  
\author{Aukkawut Ammartayakun}
\authornotemark[1]
\affiliation{
  \institution{Department of Robotics Engineering} 
  \city{Worcester Polytechnic Institute, Worcester, MA, USA}}
\email{aammartayakun@wpi.edu}
\author{Carlo Pinciroli}
\affiliation{
  \institution{Department of Robotics Engineering}
  \city{Worcester Polytechnic Institute, Worcester, MA, USA}}
\email{cpinciroli@wpi.edu}

\begin{abstract}
Ants have evolved to seek and retrieve food by leaving trails of pheromones. This mechanism has inspired several approaches to decentralized multi-robot coordination. However, in this paper, we show that pheromone trails are a fragile mechanism for coordination, and can be sabotaged to starve the colony. We introduce \emph{detractors}: malicious agents that leave a misleading, but indistinguishable, trail of food pheromone to distract and trap cooperator ants in the nest. We analyze the effectiveness of detractors with respect to parameters such as evaporation rate of misleading pheromone and fraction of detractors in the colony. In addition, we propose a countermeasure to this attack by introducing a new type of pheromone: the cautionary pheromone. Cooperator ants secrete this type of pheromone atop existing food trails as a warning. When the cautionary pheromone intensity exceeds the food pheromone intensity, cooperator ants ignore overlapping food pheromone. We show that, despite its simplicity, this defense mechanism can limit, but not nullify, the effect of detractors. Ultimately, our work shows that pheromone-based coordination, while effective, is also fragile.
\end{abstract}

\ccsdesc[500]{Computing methodologies~Distributed artificial intelligence}
\ccsdesc[500]{Computing methodologies~Multi-agent systems}
\ccsdesc[500]{Computing methodologies~Intelligent agents}
\ccsdesc[500]{Computing methodologies~Mobile agents}

\keywords{Swarm Intelligence; Collective Behavior; Adversarial Behaviors}

\begin{document}


\pagestyle{fancy}
\fancyhead{}


\maketitle 


\section{Introduction}
\label{sec:introduction}

Pheromone-based coordination is ubiquitous in nature due to its
remarkable effectiveness
\cite{wyattAnimalsChemicalWorld2014}. Stigmergy \cite{grasseReconstructionNidCoordinations1959,theraulazBriefHistoryStigmergy1999,heylighenStigmergyUniversalCoordination2016}
is a form of environment-mediated communication in which agents deposit
pheromone to coordinate.  Insect colonies such as termites, honeybees,
and ants employ stigmergy in a wide variety of tasks critical for
survival, including nest selection and construction, food collection,
and threat detection
\cite{johanbillenPheromoneCommunicationSocial1998,vonthienenPheromoneCommunicationAnts2014}.

The success of stigmergy in natural systems is an inspiration for the
design of artificial systems that must act in large, unpredictable,
and hazardous environments
\cite{heylighenStigmergyUniversalCoordination2016}. In swarm robotics,
for example, stigmergy inspired by ants trails and termite mounds has
found a wide variety of applications, including exploration and path
planning
\cite{russellAntTrailsExample1999,paytonPheromoneRobotics2001,ludwigRoboticSwarmDispersion2006},
task allocation \cite{garnierAlicePheromoneLand2007}, foraging
\cite{hamannAnalyticalSpatialModel2007,campoArtificialPheromonePath2010a},
and construction
\cite{allwrightSRoCSLeveragingStigmergy2014,wagnerSMACSymbioticMultiAgent2021}.

However, despite its effectiveness, stigmergy presents peculiar
instances of fragility, which hinder the performance of an insect
colony or even threaten its very survival
\cite{tristramd.wyattBreakingCodeIllicit2014}. One example of such
fragility is the well-known “ant mill” or “army ant syndrome,” an
emergent phenomenon where ants form a pheromone loop that traps them
into a circular motion that often ends with the collapse of the colony
\cite{schneirlaUniqueCaseCircular1944,bradyEvolutionArmyAnt2003}. In
addition, several predators and parasites evolved to utilize pheromone
to exploit the colony or its resources, sometimes with catastrophic
effects for the colony \cite{tristramd.wyattBreakingCodeIllicit2014}.

Studying the fragility of stigmergy is an important step towards
real-world deployment of robotic solutions based on this concept. For
this reason, in this paper we study an intrusion attack in which one
or more malicious agents, which we call \emph{detractors}, deposit
pheromone trails. Their explicit intent is misleading
\emph{cooperator} (benign) foraging agents and trapping them in the
nest, unable to find food. The key assumption of our work is that the
misleading pheromone, as well as the detractors that deposit it, are
indistinguishable from their benign counterpart. We show how this
attack can be performed by detractors with minimal capabilities,
comparable to those of cooperators. We assess the damage this attack
provokes with respect to parameters such as the fraction of detractors
in the colony and the evaporation time of the misleading pheromone. It
is worth noting that, while in this paper we frame the presence of
misleading pheromone trails as a deliberate attack on the colony, this
phenomenon might also arise from unintentional failures in one or more
agents — a likely occurrence for robot swarms involved in hazardous
missions.

We also study a potential countermeasure to this attack that assumes
no additional cognitive capabilities for the agents, such as enhanced
memory or learning. Specifically, we study the deposition of a new
kind of pheromone, the `cautionary' pheromone, whose relative
intensity with respect to normal pheromone offsets the probability of
a cooperator following an existing trail. We study the benefit and
limitations of this simple countermeasure with respect to key design
parameters. Ultimately, our results indicate that coordination based
on pheromone trails with minimalistic agents is effective, but also
inherently and unavoidably fragile.

The rest of this paper is structured as follows. In
Sec.~\ref{sec:related_work} we survey related work on pheromone-based
coordination in natural and artificial systems. In
Sec.~\ref{sec:breaking} we discuss how detractors attack the
colony. In Sec.~\ref{sec:fixing} we present the cautionary pheromone
and analyze its effectiveness in defending against misleading
pheromone. We conclude the paper in Sec.~\ref{sec:conclusions}.

\section{Related Work}
\label{sec:related_work}


Along with the pheromone-based coordination mechanisms that allow
insect colonies to thrive, other organisms have evolved the ability to
exploit insect pheromone to gain survival advantage. Exploitation
takes a wide variety of forms
\cite{tristramd.wyattBreakingCodeIllicit2014}, often leading to
complex evolutionary arms races between attackers and
victims. `Eavesdropping', for example, is employed by predators and
parasitoids to locate their victims. Yellowjacket wasps (\emph{Vespula
  germanica}) prey on the Mediterranean fruit-fly males
(\emph{Ceratitis capitata}) by smelling the fruit-flies’ sex
pheromone, and parasitoid wasps \emph{Telenomus euproctidis}
parasitize the eggs of their victim, the female moth \emph{Euproctis
  taiwana}, following the moth’s pheromone.

In the case of ant colonies, pheromone and chemical cues are
foundational means for coordination at large scales. This makes ant
colonies an attractive target for many `guest' species, such as
millipedes, mites, spiders, isopods, crickets, flies, butterflies,
beetles, and even snakes
\cite{bagneresChemicalDeceptionMimicry2010,bertholldoblerAnts1990,kronauerMyrmecophiles2011,rettenmeyerLargestAnimalAssociation2011}. Once
accepted in the colony, these `guest' species benefit from the abundance
of available food, protection from predators, and protection from extreme
heat and humidity offered by the nest. The main mechanisms to be
accepted in the colony include insignificance (the guest is
`invisible' to the colony), chemical camouflage (the guest `steals'
the molecules that make it smell like an ant), and chemical mimicry
(the guest produces the right molecules).

An example of attack closely related to the subject of this paper is
offered by `slave making' ants, which emit a sort of `propaganda
pheromone'
\cite{akinoChemicalStrategiesDeal2008,morganChemicalSorcerySociality2008}
interpreted by the victim ants as an alarm pheromone. The victims
instinctively respond to this pheromone with panic, making it possible
for the attackers to confuse and disperse the colony. If successful,
the attackers enter the nest and steal the pupae, which will grow to
become workers for the `slave making' ants.

Evolutionary arms races have provided social animals with
countermeasures that, partially or completely, limit the damage of
these kinds of attacks. Learning to associate danger to pheromone is
arguably one of the most effective mechanisms, but it is typically
displayed by animals with high cognitive capabilities such as mice or
hamsters. In contrast, insect colonies, and specifically ants, have been
found unable to learn to avoid their instinct to respond to pheromone
trails \cite{wenigHardLimitsCognitive2021}.

It is reasonable to conjecture that artificial systems that take
inspiration from stigmergy in insect colonies will share the fragility
of its natural counterpart. Nonetheless, this aspect has received
little attention in the literature. In particular, research in swarm
robotics has focused on showing how these mechanisms could be applied
to scenarios in which cooperation is assumed at all times, and
failures are either absent or insignificant to pheromone deposition
\cite{heylighenStigmergyUniversalCoordination2016,
  russellAntTrailsExample1999, paytonPheromoneRobotics2001,
  ludwigRoboticSwarmDispersion2006, garnierAlicePheromoneLand2007,
  hamannAnalyticalSpatialModel2007, campoArtificialPheromonePath2010a,
  allwrightSRoCSLeveragingStigmergy2014,
  wagnerSMACSymbioticMultiAgent2021,huntTestingLimitsPheromone2019}. This
is not surprising, given the technical difficulty of implementing
effective pheromone-based coordination in laboratory conditions
\cite{garnierAlicePheromoneLand2007, mayetAntbotsFeasibleVisual2010,
  arvinCOSFArtificialPheromone2015a,
  fujisawaDesigningPheromoneCommunication2014,
  russellHeatTrailsShortlived1997,
  simoninInteractiveSurfaceBioinspired2011a,
  birattariPhormicaPhotochromicPheromone2020,
  khaliqStigmergyWorkPlanning2015}. To the best of our knowledge, this
paper is the first to study how misleading pheromone affects an
artificial swarm system, and how to mitigate this issue.

\section{Attacking the Colony}
\label{sec:breaking}

\subsection{Objective}
\label{sec:breaking_problem}
\begin{figure}{t}
    \centering
    \includegraphics[width=0.5\textwidth]{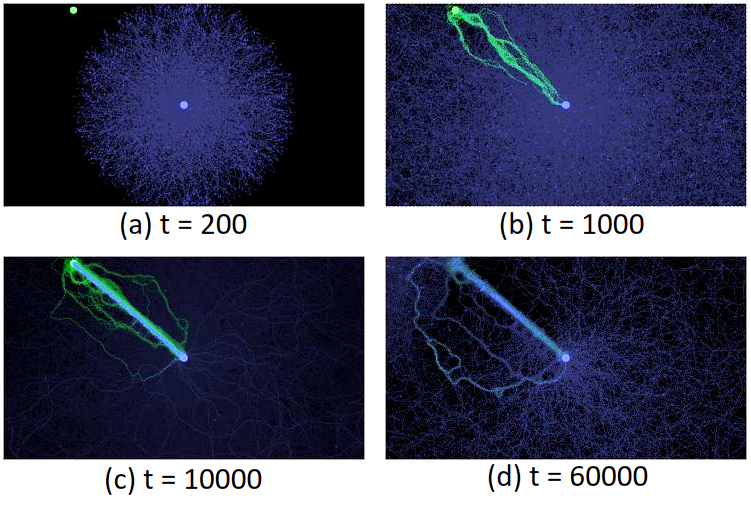}
    \caption{Typical behavior of the colony without detractors. \textmd{The home pheromone trail laid out by cooperators is shown in blue. After finding food, they lay food pheromone, shown in green. (a) Cooperators begin searching the area at time 0 and disperse radially at step 200. (b) Some ants find food and deliver it to the nest. Other cooperators encounter a trail of food pheromone and follow it to the food source. (c) Cooperators find the best path to the food. (d) Ants begin searching elsewhere after the food source is depleted.}}
    \label{fig:cooperator_timeline}
\end{figure}

We consider a scenario in which a colony of simulated ants deposits pheromone to coordinate food collection (foraging). In our model, we assume that the ants are instinctively compelled to follow trails, albeit probabilistically. The typical behavior of such a colony is shown in Figure \ref{fig:cooperator_timeline}.

To perform the attack, we introduce \emph{detractor} ants. These can be interpreted as ants whose behavior changed due to deliberate tampering (although a specific malfunction could coincidentally have the same effect). We assume detractors to be otherwise identical to benign ants (\emph{cooperators}) in terms of cognitive capabilities. Detractors lay pheromone that is intentionally meant to
mislead ants leaving the nest to find and collect food.

The main objective of this section is to formalize a simple attack
strategy for the detractors, such that the colony is unable to forage.

\subsection{Experiment Setup}
\label{sec:breaking_ant_experiment_setup}

Throughout our simulations, we fix the locations of the nest and of the food source. We simulate a colony of 1,024 ants for 50,000 steps.

\paragraph{Ant pose and pheromone storage.}
In our model, the pose of each ant (real position and angle of movement) is described by a tuple $\langle x,y,\theta \rangle$ with $(x,y) \in \mathbb{R}^2$ and $\theta \in [0, 2\pi)$. Cooperators begin at $L_\text{nest}$ with a random $\theta$. The pheromone is stored in a grid with cells of side length 4 units. Each cell stores the density of pheromone present at that location. Because the cooperators need to create paths from the nest to food and vice versa, two types of pheromone can be stored in each cell: the \emph{food pheromone}, which signals a potential trail to a food source, and the \emph{home pheromone}, for a trail home. Both values are real numbers in the range $[0,1000]$. The grid cells covered by the nest are permanently marked as nest cells. The grid cells covered by food are marked as food cells until food runs out. Cells with no pheromone nor food are marked empty. 



\paragraph{Pheromone deposition and evaporation.}
At the start of the simulation, cooperators are evenly distributed around the nest, facing outward. They begin searching for food. We call this \emph{to-food state}. As they move, ants deposit a trail of home pheromone. This pheromone is deposited with an intensity determined by
\begin{equation}
\label{eqn:breaking_intensity}
\text{Intensity} = 1000\exp{(-\lambda\tau)}
\end{equation}
where $\lambda$ is a constant and $\tau$ is the number of simulation steps since leaving the nest. At the start of the simulation, and whenever an ant reaches the nest, $\tau$ is reset to 0. If an ant successfully finds food, it switches to \emph{to-home state}. In this state, the ant secretes the food pheromone analogously to the home pheromone with the difference that $\tau$ counts the simulation steps since food was last found. Pheromone gradually evaporates after deposition, reducing in intensity linearly over time. The evaporation rate $k$ is a constant 1 unit per second. The intensity at the next timestep (Intensity') is determined by
\begin{equation}
\label{eqn:evaporation}
\text{Intensity'} = \text{Intensity}-k \Delta t
\end{equation}
If a cell already contains pheromone of a certain type and an ant tries to deposit pheromone of the same kind, the resulting intensity is the maximum between the cell intensity and the ant intensity.



\begin{table}[t]
\centering
\caption{The parameters of our ant colony model.}
\label{table:table_of_parameters}
\begin{footnotesize}
\begin{tabular}{ p{0.08\textwidth} p{0.04\textwidth} p{0.3\textwidth} } 
 \hline
 \textbf{Symbol} & \textbf{Value} & \textbf{Description} \\  \hline
 $n$ & 1024 & Number of ants in simulation \\ 
 $N$ & 50,000 & Number of simulation steps per experiment  \\ 
 $W$ & 1920 & Simulation world width  \\ 
 $H$ & 1080 & Simulation world height \\ 
 $c$ & 4 & Grid cell size \\ 
 $L_\text{food}$ & (372,36) & Location of food source \\
 $r_\text{food}$ & 16 & Radius of food source \\
 $L_\text{nest}$ & (960,540) & Location of nest \\
 $r_\text{nest}$ & 20 & Radius of nest \\
 $v$ & 50 & Ant movement per second \\ 
 $\Delta t$ & 0.016 & Time per simulation step \\ 
 $\theta^s_\text{max}$ & $0.8\pi$ & Range of vector generation for direction selection \\ 
 $l^s_\text{max}$ & 40 & Maximum magnitude of vector generation for direction selection \\ 
 $\eta$ & $0.1\pi$ & Coefficient for random noise \\
 $\lambda$ & 0.01 & Coefficient for pheromone intensity \\ 
 $\tau_{\text{turn}}$ & 7 & Simulation steps before turning \\
 $\tau_{\text{attack}}$& 100 & Simulation steps after which detractors begin to secrete misleading pheromone \\
 $\chi$ & 32 & Number of random vectors generated for direction selection \\  \hline
\end{tabular}
\end{footnotesize}
\end{table}

\begin{figure}
    \centering
    \includegraphics[width=0.285\textwidth]{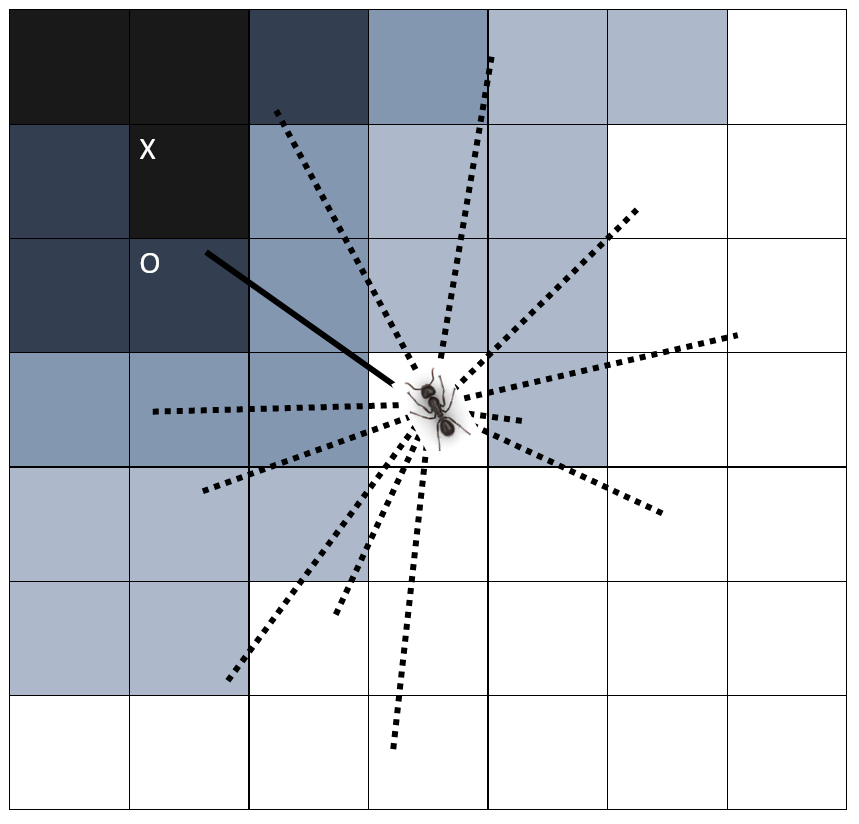}
    \caption{Ant's motion. \textmd{Pheromones are stored in the cells. The shade of a cell indicates the intensity of the pheromone, with darker shades signifying higher intensity.}}
    \label{fig:direction_decision}
\end{figure}

\paragraph{Ant motion.}
At each simulation step $\tau_{\text{turn}}$, each ant generates $\chi$ random `probing' vectors, with length chosen uniformly in $[0,l^{s}_{\text{max}}]$ and angle chosen uniformly in $[-\theta^{s}_{\text{max}}, \theta^{s}_{\text{max}}]$. Each vector starts at the ant and ends in a cell, as shown in Figure \ref{fig:direction_decision}. The cell with the highest pheromone intensity is picked, and the direction to that cell is denoted as $\theta^n$. An example is shown in Figure \ref{fig:direction_decision}, where the vector shown as a solid line leads to the cell with the highest pheromone intensity (marked \textbf{O}). Another cell within range (marked \textbf{X}) has a higher intensity than \textbf{O}, but in our example, none of the randomly generated vectors point to it, so it is not considered. If no grid cell in the vicinity contains pheromone, the direction defaults to straight. Once the direction $\theta^n$ is chosen, uniformly distributed noise $w \sim \mathcal{U}(-\eta,\eta)$ is added to it. The next pose for ant $\langle x', y', \theta' \rangle$ is determined by:
\begin{equation}
\begin{aligned}
\label{eqn:ant_next_position}
x' &= x + v \cos(\theta^{n}+w) \Delta t\\
y' &= y + v \sin(\theta^{n}+w) \Delta t\\
\theta' &= \theta^n + w
\end{aligned}
\end{equation}
in which $v$ is the (constant) speed of the ant and $\Delta t$ is the time per simulation step.
During motion, if an ant encounters a boundary, food cell, or nest cell, a `law of reflection' logic is applied to reverse the direction of movement. 
The experiments are based on a simulator hosted at \url{https://github.com/NESTLab/AntSimulator}.

\subsection{Approach}
\label{sec:breaking_approach}

\paragraph{Main idea.}
The objective for the detractors is to mislead or distract the cooperators from finding the food source. In the quest for a minimalistic but effective approach to obtain this result, we hypothesized that a simple strategy for detractors is to constantly secrete food pheromone. The secretion intensity follows \eqref{eqn:breaking_intensity}, with $\tau$ calculated as the number of simulation steps since secretion started. In the rest of this paper, we will denote this with $\tau_d$.

\paragraph{Detractors' pheromone deposition.}
We hypothesize that a high concentration of misleading pheromone increases the likelihood of a successful attack. Early trials showed interesting results when we pointed detractors in the direction of the food source, so we chose to give them this potential advantage. Detractors all start in the center of the nest, facing the food source. They stay dormant at the start of an experiment, and after $\tau_{\text{attack}}$ simulation steps begin to secrete misleading food pheromone. Cooperators then trust and follow these highly concentrated food trails.

\begin{figure}
    \centering
    \includegraphics[width=0.5\textwidth]{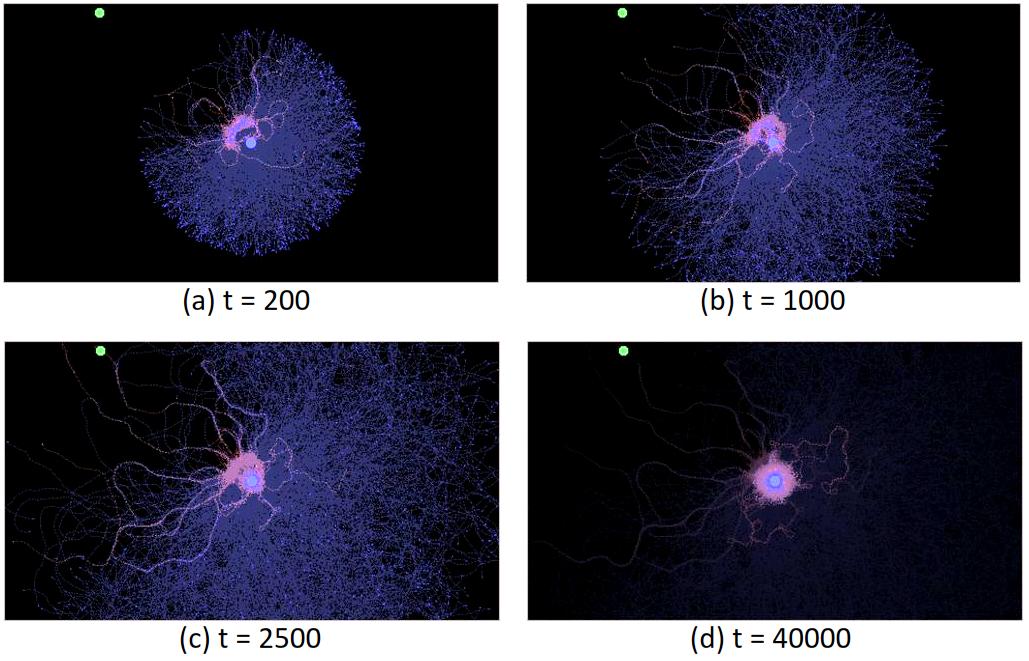}
    \caption{Visualization of detractors' impact on foraging task over time. \textmd{Home pheromone is shown in blue and misleading pheromone in red. Other food pheromone would be shown in green, but is not present.}}
    \label{fig:detractor_wall}
\end{figure}

\paragraph{Basic misleading pheromone effect.}
Both detractors and cooperators follow food pheromone trails. Neither can distinguish misleading trails from trustworthy ones. As a result, detractors continue to reinforce each others' misleading trails, building and then strengthening a blockade (as seen in Figure \ref{fig:detractor_wall} (a)). Because of the focused attack, any cooperators heading in the direction of food will be distracted by the ruse. Cooperators exploring other areas will not find food regardless, and as such, any attack there would be a waste of detractors' limited numbers. This approach works for roughly 10,000 simulation steps, but it has a critical flaw: $\tau_d$ continually increases, driving intensity down. Thus, the trap laid by the detractors grows weaker over time, and ultimately fails to fool cooperators.

\paragraph{Improving the duration of misleading pheromone.}
To address the issue of misleading pheromone weakening over time, we introduce a `refill' logic, whereby detractors' $\tau_d$ resets to 0 each time they happen to visit the nest. In Figure \ref{fig:detractor_wall} (b) we see an example of detractors finding and traveling to the nest. Subsequently, in Figure \ref{fig:detractor_wall} (c) detractors grow concentrated around the nest, resetting their $\tau_d$. With the nest completely enveloped by misleading pheromone with maximum intensity, cooperators struggle to escape. Figure \ref{fig:detractor_wall} (d) shows that cooperators, unable to escape, eventually see their home pheromone fully evaporated.

\subsection{Evaluation}
\label{sec:breaking_evaluation}
To evaluate the performance of the detractors, we consider two parameters. 
The first is evaporation rate $k$ from \eqref{eqn:evaporation}. The second is the fraction of detractors in the colony. We evaluated the success of cooperators using two metrics: \emph{(i)} the amount of food collected and delivered per cooperator, and \emph{(ii)} the fraction of colony that was able to collect and deliver food. When a cooperator discovers and reaches food, it collects a food bit, which is considered delivered when it is brought to the nest. Each cooperator can travel with only one food bit.

\begin{figure}
\includegraphics[width=\linewidth]{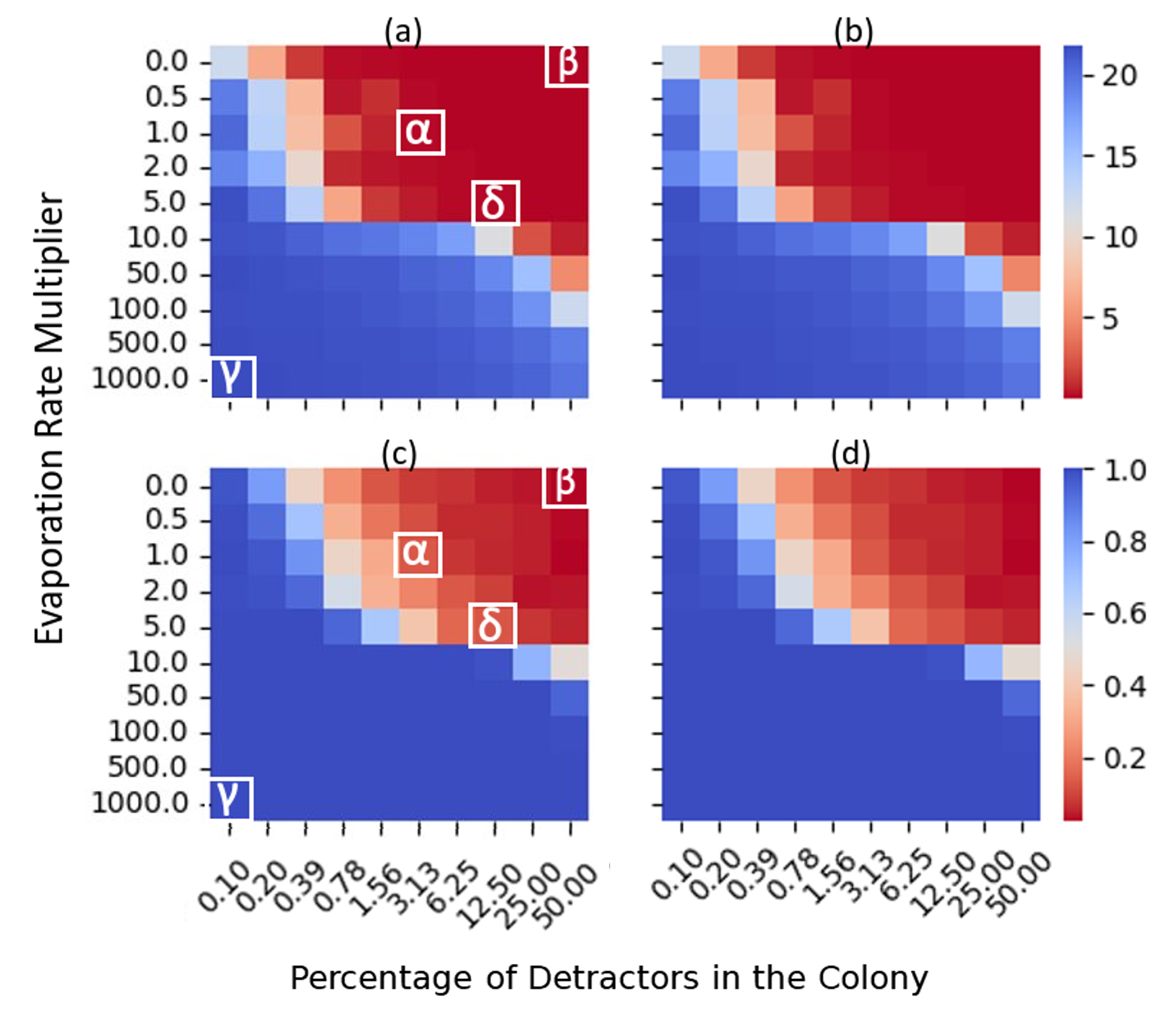}
\caption{Heatmap of cooperator success.
    \textmd{
    Heatmaps (a) and (b) show the average food bits collected and delivered per cooperator, respectively. Heatmaps (c) and (d) show the fraction of cooperators that collected and delivered food, respectively. Each data point in the heatmap is the average of 20 simulation runs. The configurations denoted by $\alpha, \beta, \gamma$, and $\delta$ are chosen for extended evaluation in Figure \ref{fig:breaking_evaluation}.}}
\label{fig:breaking_heatmap}
\end{figure}
\begin{figure}
    \includegraphics[width=\linewidth]{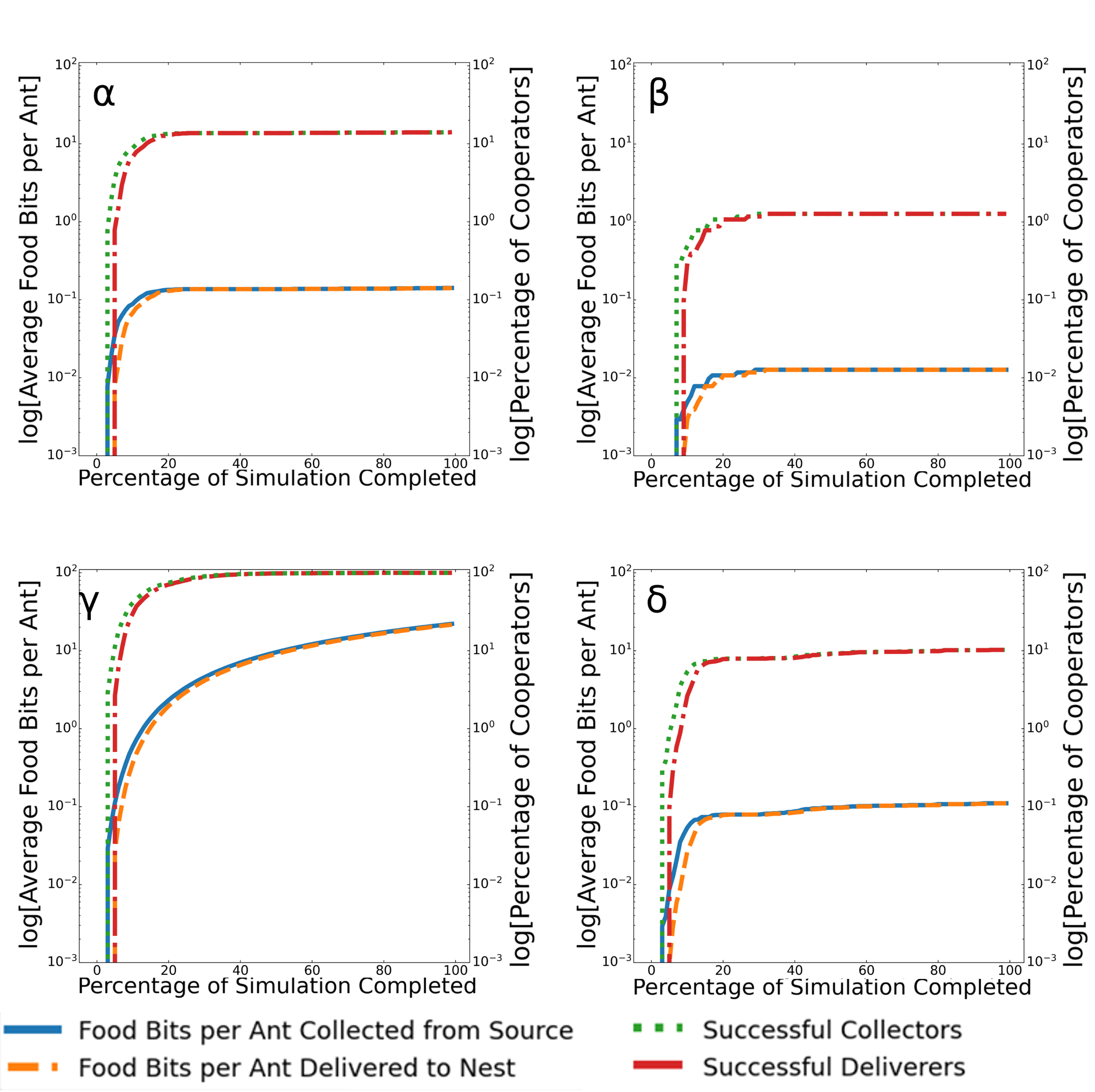}
\caption{Line graph of cooperator success over time.
\textmd{Graphs $\alpha - \delta$ show the progression of two metrics over the course of four noteworthy configurations taken from Figure \ref{fig:breaking_heatmap}. Semi-logarithmic scale is used for comparisons.}}
\label{fig:breaking_evaluation}
\end{figure}

In Figure \ref{fig:breaking_heatmap}, we compare the results of different configurations where we vary the evaporation rate and the proportion of detractor ants in the colony. On the y axis, we alter the evaporation rate with a multiplying factor against the standard evaporation rate used by cooperators. At 1000, misleading pheromone evaporates at 1000 times the rate of other pheromones (it is very weak). At 0, it never evaporates (and is extremely strong). Red data points signify results that favor detractors (food collection was negatively impacted), while blue data points signify results that favor cooperators (food collection was typical). The heatmap is generated by taking the average performance over 20 simulations for each configuration. 

The heatmap reveals a boundary between cooperator and detractor success. Decreasing the evaporation multiplier (i.e., making misleading pheromone stronger) has a similar effect to increasing the relative population of detractors. Both hamper the colony's ability to collect food. Armed with a pheromone that never evaporates, detractors that make up a mere 0.39\% of the colony are enough to disrupt it. The colony's food collection per cooperator in this configuration is a meager 5\% of what could be collected in the absence of detractors. Even with weaker misleading pheromone, a makeup of 1\% of detractors can disrupt the colony. However, an evaporation rate greater than about 5 times that of normal food pheromone requires a significant fraction of detractors to cause any disruption.

Heatmaps (a) and (b) in Figure \ref{fig:breaking_heatmap} report the total food collected by cooperators and are nearly identical. This indicates that cooperators which find food reliably deliver it to the nest. This can be explained by the fact that ants in \emph{to-home state} rely on the home pheromone, not the food pheromone, and our detractors do not interfere with the home pheromone. This is confirmed by the graphs in Figure \ref{fig:breaking_evaluation} in which the food collection and delivery lines are almost identical. This suggests that cooperators do not get lost on their way back to the nest once they have collected a food bit.

Heatmaps (c) and (d) in Figure \ref{fig:breaking_heatmap} show the number of successful foragers and generally confirm the above discussion. However, these heatmaps show a comparatively smaller area of detractor dominance with respect to heatmaps (a) and (b). This suggests that altering the evaporation rate multiplier and the relative population of detractors has a greater impact on food collection as measured by the contribution of individual cooperators. In other words, it is easier for detractors to decrease the efficiency of food collection than the involvement of individual cooperators.

To further analyze the impact of the evaporation rate and the fraction of detractors, we highlight four noteworthy configurations from Figure \ref{fig:breaking_heatmap} (a): $\alpha, \beta, \gamma$ and $\delta$. In configuration $\gamma$, detractors are nearly absent, and those that do exist have a negligibly present misleading pheromone, making them virtually harmless. The fraction of ants that successfully deliver food (shown in red dash-dot pattern in Figure \ref{fig:breaking_evaluation} ($\gamma$)) approaches 100\%. 60\% through the simulation, almost all the ants had found the food source. The trend of food delivered to the nest (shown in an orange dashed pattern in Figure \ref{fig:breaking_evaluation} ($\gamma$)) suggests linear growth, as cooperators steadily transport food once discovered.

In Figure \ref{fig:breaking_evaluation} ($\beta$), detractors comprise half the colony, and misleading pheromone never evaporates. This is an extreme scenario where cooperators invariably fail. To shed light on the dynamics of this scenario, we compare the fraction of cooperators that manage to bring home at least one food bit with the average amount of food collected by the colony. The results show that these metrics are almost identical: 0.022363 and 0.022461 respectively. Only 2.23\% of the ants were able to collect and deliver food throughout the simulation run. However, after delivering that first food bit, these ants never found food again. A maximum of 0.0098\% of the ants were able to find food a second time.

Figure \ref{fig:breaking_evaluation} ($\alpha$) and ($\delta$) depict moderate scenarios. We chose configuration $\alpha$ such that the minimum few detractors could have a large effect on food delivery when the evaporation rate of misleading pheromone is kept the same as that of trustworthy food pheromone. Cooperators' performance improves when the misleading evaporation rate multiplier increases from 5 to 10 (becomes weaker). As such, we study the behavior of misleading pheromone when the evaporation rate is 5 times that of the trustworthy food pheromone. In this scenario ($\delta$), we increase the population of detractors by a factor of 4.


The evaporation rate of misleading pheromone in Figure \ref{fig:breaking_evaluation} ($\alpha$) is analogous to the cooperator ants' food pheromone, and the fraction of detractors is 3.13\% of the total population. Even with such a small fraction of detractors, only 13.10\% of the cooperators could collect and deliver food. This value is reached quickly: only 20\% into the simulation. We again see in Figure \ref{fig:breaking_evaluation} ($\alpha$) that once a cooperator delivers food to the nest, it usually does not return to the food source. Most of the ants that were not misled by detractors and found food during the first 20\% of the simulation were trapped by misleading pheromone upon returning to the nest. The final average of food bits per cooperator is only 0.14, compared to 21.74 in the absence of detractors. This highlights the effect of even a small presence of well-motivated detractors.

In Figure \ref{fig:breaking_evaluation} ($\delta$), 12.5\% of the ants are detractors, but the evaporation rate of misleading pheromone is 5 times faster than the trustworthy one. The cooperators were able to collect only about 0.12 food bits per ant on average during the simulation. The increasing trend of average food bits per ant indicates that cooperators were still able to find the food source, albeit very slowly.

\section{Defending the Colony}
\label{sec:fixing}

\subsection{Problem Statement}
\label{sec:fixing_problem}


The discussion in Section \ref{sec:breaking_evaluation} showcases the effectiveness of misleading pheromone in hindering the colony's ability to collect food. We note, however, a pattern in the detractors' success. They form a \emph{small} region where they can reinforce misleading pheromone that builds to a strong intensity despite their minimal numbers. Once this small, but powerful, barricade is formed, cooperators circle that area endlessly in a futile search for food.

\subsection{Approach}
\label{sec:fixing_approach}

\paragraph{Main ideas.}
Cooperators benefit from their large numbers, but they might further benefit from an additional tool to overcome misleading pheromone. This new tool, however, must be difficult for detractors to exploit. Our approach is to employ a third type of pheromone: a `cautionary' pheromone. This pheromone, laid by cooperators, serves to caution others of a false food trail. We further add a new parameter: \emph{patience}. Patience is modeled as a counter that limits the amount of time steps ants are willing to follow a food trail. Cooperators in \emph{to-food state} continually secrete cautionary pheromone with intensity given by
\begin{equation}
\label{eqn:fixing_intensity}
\text{Intensity} = 1000\exp{(-\lambda\rho)}
\end{equation}
where patience is denoted with $\rho$ and $\rho \in [0, \rho_\text{max}]$. Patience is initialized with $\rho_\text{max}$. When a cooperator starts out in search of food, the cautionary pheromone secreted is weak. As time goes on, patience decreases and cautionary pheromone increases in intensity. When an ant finds food, its patience resets to $\rho_\text{max}$. Because the cautionary pheromone is only meant to counter misleading trails, patience increases when there is no detectable food pheromone (there is nothing against which to caution). The patience increments are such that it can refill from 0 to $\rho_{\text{max}}$ in $t_p$ amount of timesteps. The evaporation rate is set to 1 unit per second.

\begin{figure}[t]
    \centering
    \includegraphics[width=0.3\textwidth]{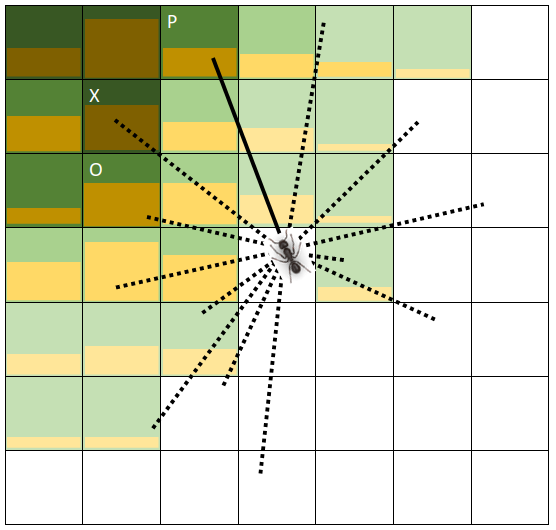}
    \caption{Ant's decision on direction of movement in the presence of cautionary pheromone.
        \textmd{Ratio of yellow to green indicates ratio of cautionary pheromone to food pheromone. Darkness indicates pheromone intensity.}}
    \label{fig:counter_decision_making}
\end{figure}

\paragraph{Cooperator motion.}
The overall motion mechanism remains the same as of Section \ref{sec:breaking_ant_experiment_setup}, with the introduction of one more pheromone to consider. Multiple random vectors are generated to determine the next direction, as shown in Figure \ref{fig:counter_decision_making}. In this figure, the three cells to observe are denoted by \textbf{O}, \textbf{X}, and \textbf{P}. A random vector points to cell \textbf{X} which has the highest pheromone intensity in the sensing range. However, the cautionary pheromone intensity in this cell exceeds the intensity of the food pheromone. Therefore, this cell is ignored. Next, cell \textbf{O} is considered, but also here the intensity of the cautionary pheromone exceeds that of the food pheromone. Finally, in cell \textbf{P}, the intensity of the food pheromone is higher than the cautionary pheromone's, so \textbf{P} is chosen as the next direction. If the cooperators are stuck in the misleading pheromone's trap, eventually all the cells with food pheromone are overpowered by cautionary pheromone. Once all the cells surrounding the ant cannot be considered for direction, the ant defaults to the straight direction. In the absence of any direction other than straight, noise $w$ in \eqref{eqn:ant_next_position} will be the major driver to determine the ant's next pose. 


\begin{figure}[t]
    \centering
    \includegraphics[width=0.5\textwidth]{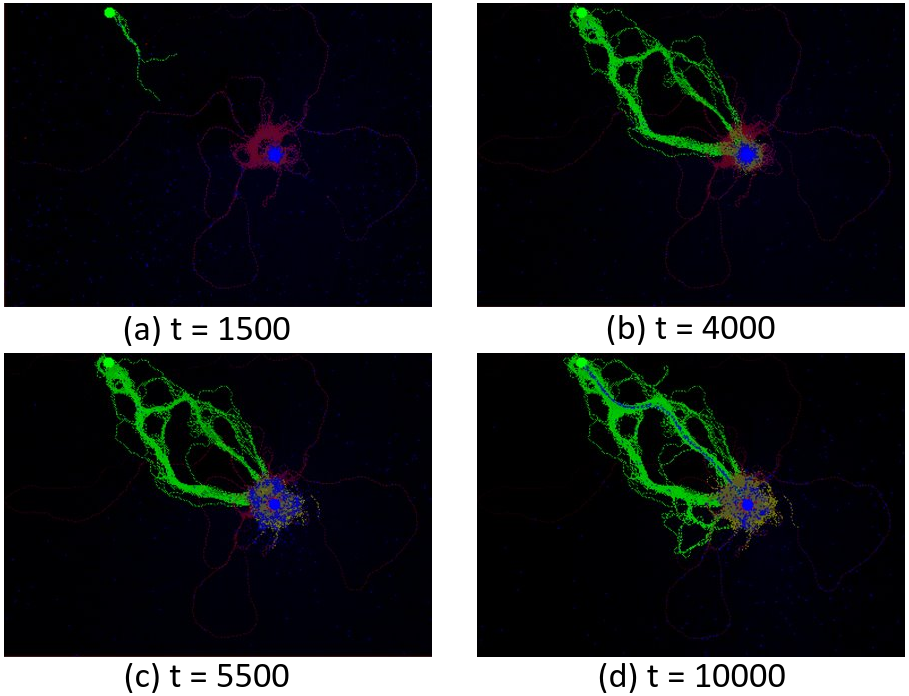}
    \caption{Cooperators using cautionary pheromone to counter the detractors' misleading pheromone. \textmd{Misleading pheromone is shown in red, food pheromone is shown in green, and cautionary pheromone is shown in yellow.}}
    \label{fig:counter_pheromone_timeline}
\end{figure}

In Figure \ref{fig:counter_pheromone_timeline}, we see cooperators successfully leverage cautionary pheromone (shown in yellow) to overcome the trap of misleading pheromone (shown in red). The detractors' attack in (a) begins as in Figure \ref{fig:detractor_wall}, but Figure \ref{fig:counter_pheromone_timeline} (b) shows a hint of yellow gradually covering the red. This visualizes cooperators gradually laying cautionary pheromone (in increasing intensity) as they follow the trail of food pheromone (which in this case is misleading). As the misleading pheromone trap draws cooperators, they reinforce each other's cautionary pheromone trails until they all lose patience, cautionary pheromone intensity exceeds food pheromone intensity, and ants ultimately ignore the misleading pheromone.

Once misleading pheromone can be ignored, cooperators search for and follow alternative food trails. Figure \ref{fig:counter_pheromone_timeline} (c) shows cooperators vacating the misleading pheromone trap around the nest: more with every time step. Finally, in Figure \ref{fig:counter_pheromone_timeline} (d), cooperators identify a legitimate food source and lay food pheromone that overpowers misleading pheromone. Cooperators in \emph{to-home state} do not secrete cautionary pheromone, and those in \emph{to-food state} find food so quickly that patience remains high and cautionary pheromone weak. Once the delivery of food steadies to a relatively normal (as if misleading pheromone did not exist), we consider cooperators successful in having mounted a defense against the attack.


    
    
    
    
    


\subsection{Evaluation}
\label{sec:fixing_evaluation}

We use the same method and metrics as in Section \ref{sec:breaking_evaluation} to evaluate the effectiveness of the cautionary pheromone. We study the effect of maximum patience ($\rho_\text{max}$) and refill rate of patience ($t_p$) as variables to alter the potency of the cautionary pheromone.

\begin{figure}[t]
    \centering
    \includegraphics[width=0.5\textwidth]{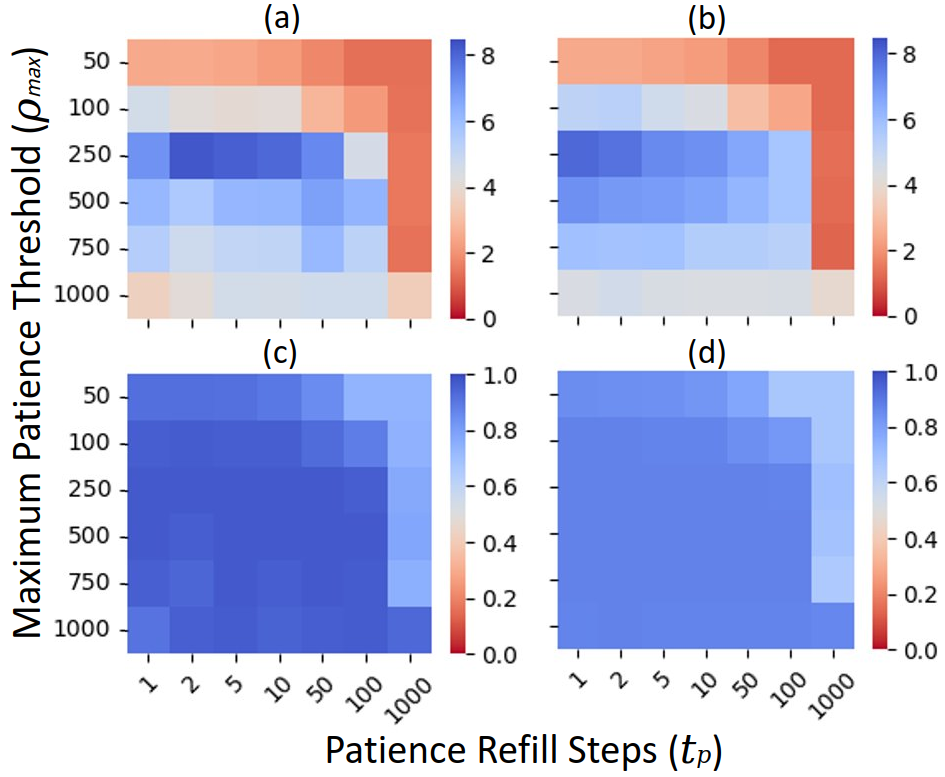}
    \caption{Heatmap of cooperator ants' success against the detractors' attack. \textmd{Heatmaps (a) and (b) show the food bits per ant collected by the cooperator ants at the end of the simulation according to the cases in Figure \ref{fig:breaking_evaluation} ($\alpha$) and ($\delta$) respectively. (c) and (d) show the fraction of cooperators that were able to collect food in the experiments (a) and (b), respectively. As  discussed in Section \ref{sec:breaking_evaluation}, detractors were unable to distract cooperators returning to the nest. Therefore, we do not discuss the effect of cooperators' ability to return to the nest. Each configuration in the heatmap represents the average result of 20 simulations.}}
    \label{fig:fixing_graph}
\end{figure}

The heatmaps in Figure \ref{fig:fixing_graph} (a) show the effect of different configurations of patience. Experiment setups for heatmaps (a) and (c) show a colony with 3.13\% detractors and a misleading pheromone evaporation rate multiplier of 1. This configuration corresponds to the case in Figure \ref{fig:breaking_evaluation} ($\alpha$). Similarly, heatmaps (b) and (d) correspond to the case in Figure \ref{fig:breaking_evaluation} ($\delta$), with 12.5\% of the colony population as detractors with an evaporation rate multiplier of 5. 

In Figure \ref{fig:fixing_graph} (a) we see that cautionary pheromone has less impact when maximum patience has a very low or very high value. Similarly, when patience refill rate is too slow (steps to reset patience are high), cooperators struggle to find food. In both cases, cooperators become too cautious and question even trustworthy food trails, finding it difficult to collect food reliably. Even in these scenarios, we see that the food bits collected per ant is in the range 1-5. This is better performance than without cautionary pheromone, where cooperators collected an average of 0.14 food bits per ant. We observe the best results when the maximum patience is set to 250 and $t_p$ in the range of 2-10. In these three configurations, cooperators are able to collect more than 8 food bits per ant, with more than 96\% of the cooperators collecting at least one food bit. In the corresponding case with no cautionary pheromone (Figure \ref{fig:breaking_evaluation}($\alpha$)), only 13.1\% of the cooperators collected food, and only 0.14 food bits per ant. Utilizing cautionary pheromone improves cooperator contribution 7 fold and food collection by a magnitude of 58.

When cautionary pheromone is applied to the scenario in Figure \ref{fig:breaking_evaluation}($\delta$), the results resemble those of configuration (a) in Figure \ref{fig:fixing_graph}. Referencing Figure \ref{fig:fixing_graph} (b) and (d), we observe that cooperators struggle to collect food when $\rho_\text{max}$ is very low or $t_p$ is very high. When maximum patience is set to 250 and $t_p$ set to 1-2 steps, 87\% of the cooperators collected cumulatively almost 8 food bits per ant.


Cautionary pheromone with a maximum patience of 250 and $t_p$ set to 1
 completely fails in the extreme scenario of Figure \ref{fig:breaking_evaluation} ($\beta$), where detractors comprise 50\% of the population and misleading pheromone never evaporates.
\begin{figure*}[ht]
    \centering
    \includegraphics[width=0.4\textwidth]{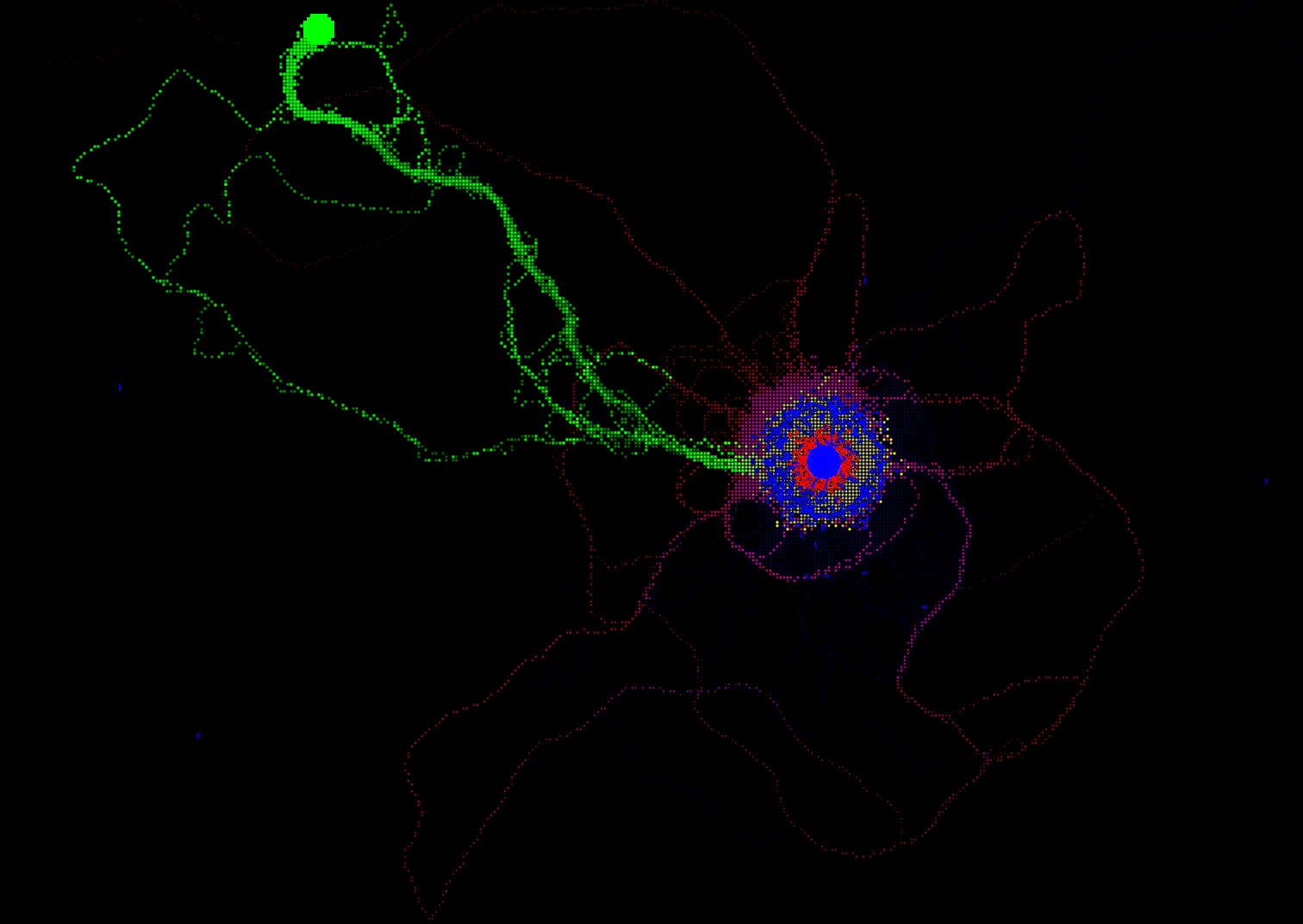}
    \includegraphics[width=0.5\textwidth]{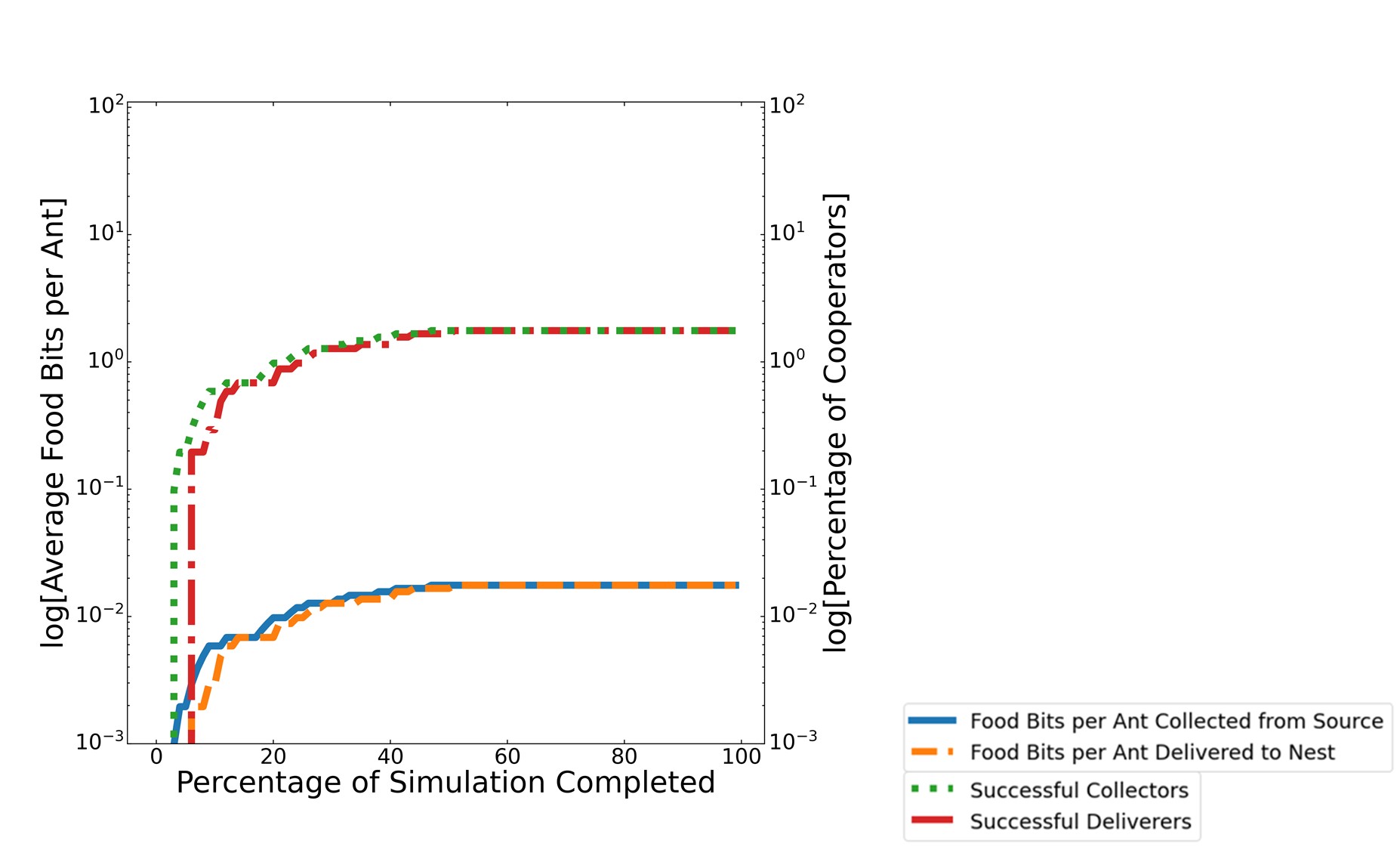}
    \caption{Cooperators trying to defend against a population of 50\% detractors whose misleading pheromone never evaporates (t=20,000 simulation steps). \textmd{Food pheromone is shown in green, misleading pheromone in red, cautionary pheromone in yellow, and cooperators in blue.}}
    \label{fig:fixing_max_detractors}
\end{figure*}
It is important to note that in this scenario, misleading pheromone never evaporates, but cautionary pheromone does. The simulations reveal that the cooperators reached an equilibrium state in which they created a circular formation around the nest (as shown Figure \ref{fig:fixing_max_detractors}). The freshly laid cautionary pheromone almost immediately evaporated and could not overpower the misleading pheromone. Cooperators were unable to break out of this equilibrium state and only 0.35 food bits per ant could be collected. In addition, the food bits collected and the fraction of cooperators that collected food resembled to the case with no cautionary pheromone, Figure \ref{fig:breaking_evaluation} ($\beta$). In other words, few ants could find food a second time.

\section{Conclusions}
\label{sec:conclusions}
In this paper, we show how stigmergy can be exploited to disrupt a colony's foraging task. We show how detractors that leave misleading trails of food pheromone distract and trap cooperator ants near the nest. Our findings show that as few as 3\% of detractors in the colony can reduce food collection 150 fold. We observed similar results when increasing the population of detractors but decreasing their pheromone potency. Even with a pheromone that evaporates five times as fast as cooperators, detractors can still disrupt a colony.

We implemented a countermeasure against misleading pheromone: cautionary pheromone. Cooperators use this pheromone to warn others of questionable food trails when they do not find food. Our results showed that cooperators leveraging this defense mechanism, in some cases, could improve their performance by a factor of 57 in the face of an attack. However, while effective in moderate cases, cautionary pheromone could still fail to thwart a large population of detractors. Pheromone is a powerful coordination mechanism, but it is fragile.

Future work focuses on discovering new types of attacks and possible countermeasures that involve learning on both sides.



\begin{acks}
  The authors acknowledge support from NSF grant \#1939061.
\end{acks}



\bibliographystyle{ACM-Reference-Format} 
\bibliography{refs}


\end{document}